\documentclass[3p,times,procedia]{elsarticle}
\usepackage{nupha_ecrc}

\volume{00}

\firstpage{1}

\journalname{Nuclear Physics A}

\runauth{Aleksi Vuorinen}

\jid{nupha}

\jnltitlelogo{Nuclear Physics A}

\usepackage{amssymb}
\usepackage{amsmath}

\usepackage[figuresright]{rotating}

\begin{document}

\begin{frontmatter}



\dochead{XXVIIth International Conference on Ultrarelativistic Nucleus-Nucleus Collisions\\ (Quark Matter 2018)}

\title{Neutron stars and stellar mergers as a laboratory for dense QCD matter}


\author{Aleksi Vuorinen}

\address{Department of Physics and Helsinki Institute of Physics, P.O.~Box 64, FI-00014 University of Helsinki, Finland}

\begin{abstract}
Neutron star observations, including direct mass and radius measurements as well as the analysis of gravitational wave signals emitted by stellar mergers, provide valuable and unique insights into the properties of strongly interacting matter at high densities. In this proceedings contribution, I review recent efforts to systematically constrain the equation of state (EoS) of dense nuclear and quark matter using a combination of ab initio particle and nuclear physics calculations and astrophysical data. In particular, I discuss the constraints that the gravitational wave observation GW170817 has placed on the EoS, and comment on the future prospects of improving the accuracy, to which this quantity is known.
\end{abstract}


\end{frontmatter}



\section{Introduction}
\label{intro}

Neutron star (NS) cores are believed to contain QCD matter in a range of densities that extends far beyond what can be probed in any terrestrial experiment \cite{Lattimer:2004pg,Brambilla:2014jmp}. Recalling that the General Relativistic description of the hydrostatic equilibrium present inside the stars offers an essentially 1-1 mapping between the equation of state (EoS) and the mass-radius (MR) relation of the NSs, it becomes clear that any astrophysical measurement that constrains the macroscopic properties of these extremely compact objects has the potential to improve our understanding of the way nature works at subnuclear scales. While these basic facts have been understood for a long time, it is only recently that stellar astrophysics has become a fully functioning laboratory of QCD matter. This is due on one hand to the rapidly improving precision of NS mass and radius measurements (see e.g.~\cite{Demorest:2010bx,Antoniadis:2013pzd,Yamamoto:2017wre,Guillot:2013wu,Nattila:2015jra,Bogdanov:2016nle,Nattila:2017wtj,Steiner:2017vmg}) and on the other hand to the recent first-ever observation of a binary NS merger. Remarkably, the latter was recorded in two distinct and highly complementary ways: through gravitational waves (GWs) detected by the LIGO and Virgo observatories \cite{TheLIGOScientific:2017qsa,Abbott:2017dke} and via an electromagnetic signal seen by the Fermi telescope and many other instruments \cite{GBM:2017lvd,Monitor:2017mdv,Soares-Santos:2017lru,Hallinan:2017woc}. 

Besides marking the birth of a long-awaited era of multimessenger astronomy, the NS merger observed in the fall of 2017 immediately set stringent limits on the properties of NSs. These new constraints came in the form of upper limits placed on the tidal deformabilities of the colliding stars, a quantity that parameterizes the extent they affected each others' gravitational fields in the inspiral phase of the merger (see e.g.~\cite{Hinderer:2009ca} for a detailed discussion of the effect). In short, the magnitude of tidal effects correlates with the physical extent of the stars, and conversely the absence of such an imprint in the observed GW signal can be used to set an upper limit for the stellar radii and thereby also the stiffness of the matter the stars are composed of. Future measurements of the inspiral phase of NS mergers are expected to further tighten these limits (see also the recent LIGO/Virgo update \cite{Abbott:2018exr}), and an eventual recording of a post-merger ``ringdown'' signal, characterizing the relaxation of the system towards its final state, is hoped to provide direct information on the NS EoS even away from the limit of very low temperatures \cite{Baiotti:2016qnr,Most:2018eaw}.

On the microphysics side, vigorous research clearly needs to be carried out in several independent directions. Firstly, the new observational data on NSs needs to be converted to model-independent constraints for the properties of dense QCD matter. This topic has already attracted significant attention in the field and is also the main topic of these proceedings (cf.~references in sec.~4 below). Alongside these efforts, it is, however, also important to further develop the theoretical methods, with which dense nuclear and quark matter are described, so that proper physical conclusions may be drawn from the form of the constrained EoS. Of particular importance here are efforts, carried out equally on the nuclear and particle physics sides, to bring the first-principles descriptions of confined nuclear matter and deconfined quark matter closer to each other (cf.~references in sec.~2 below). The ultimate hope is that equipped with improving observational limits on NS properties and results from ab initio calculations on the particle and nuclear physics sides, one will eventually be able to draw firm conclusions about the nature of the matter inside NS cores. On the astrophysics side, the key question is naturally, whether deconfined quark matter exists inside NSs, or whether it might be produced in NS mergers \cite{Most:2018eaw}. At the same time, on the microphysics side the goal is to map the cold and dense regime of the QCD phase diagram, i.e.~help discover the location and properties of a possible first-order phase transition line between the confined and deconfined phases that would end at a tricritical point \cite{Stephanov:2004wx}. Clearly, these two goals are closely related to each other, and the success of one is intimately tied to the other.

In the conference proceedings at hand, I review the current status of efforts to pin down the NS matter EoS in a model-independent way, using a combination of ab initio nuclear and particle theory calculations and astrophysical observations. Here, I follow a series of recent papers \cite{Hebeler:2013nza,Kurkela:2014vha,Fraga:2015xha,Gorda:2016uag,Annala:2017llu,Most:2018hfd}, concentrating in particular on what we have learned from the first-ever NS merger event, discussed in \cite{Annala:2017llu,Most:2018hfd}. The results I present will predominantly be from \cite{Annala:2017llu}. An important assumption in them is that the NSs, whose masses or radii have been inferred in different ways, are all main sequence stars; for alternative scenarios, such as twin stars, the reader is adviced to consult e.g.~\cite{Alvarez-Castillo:2017qki,Alvarez-Castillo:2018pve} and references therein.

This article is structured as follows. In Section 2, I first briefly walk the reader through state-of-the-art results of the EoS of sub-saturation-density nuclear matter and high-density quark matter, respectively, and in Section 3 explain how one can build interpolating NS matter EoSs that respect these limits. In Section 4, I then discuss the effects of astrophysical measurements on the uncertainty of the interpolating EoS cloud, covering in turn direct mass and radius measurements as well as the recent GW data of LIGO and Virgo. Finally, in Section 5 I provide a brief outlook on the future prospects of the field, commenting on the observational and theoretical inputs needed to accurately pin down the NS matter EoS and address questions such as the possible existence of deconfined quark matter inside NSs.

\section{Microscopic limits on the EoS}
\label{limits}

In the region around and below nuclear matter saturation density $n_s$, the natural starting point in the determination of the NS matter EoS is clearly not the Lagrangian of full QCD. Rather, one should consider a low-energy Effective Field Theory (EFT) description of the system in terms of nucleons and their effective interactions. Relying solely on controlled first principles (i.e.~ab initio) methods, the standard procedure is to merge together a tightly constrained EoS for the outer crust of the NS \cite{Chamel:2008ca} to a nuclear matter EoS derived from Chiral Effective Theory (CET), extending somewhat beyond nuclear matter saturation density \cite{Tews:2012fj,Drischler:2016djf}. Some very promising progress has been achieved in generalizing these results to higher densities using Functional Renormalization Group techniques \cite{Drews:2016wpi}, but at the moment a consensus between different independent calculations only exists up to the saturation regime. This is clearly an unsatisfactory situation, considering that the cores of NSs may easily contain matter up to more than 5 times the saturation density.

At the opposite end of extremely high (energy) densities, QCD is known to become asymptotically free, thus facilitating a weak coupling expansion of the EoS of deconfined quark matter. Such an approach is, however, known to be plagued by many issues, ranging from the appearance of physical infrared divergences (see e.g.~\cite{Mogliacci:2013mca,Laine:2016hma} for discussions of this issue in the highly analogous case of high temperatures) to the fact that quark pairing is known to affect the ground state of QCD in the limit of high densities and small temperatures \cite{Alford:2007xm}. At the moment, the state-of-the-art perturbative EoS at exactly zero temperature originates from a three-loop, or ${\mathcal O}(g^4)$, calculation including non-zero quark masses \cite{Kurkela:2009gj} (see also \cite{Freedman:1976ub,Vuorinen:2003fs} for the corresponding result in the zero-mass limit), to which the first genuinely new perturbative order of ${\mathcal O}(g^6\ln^2 g)$ has been very recently determined \cite{Gorda:2018gpy,new2}. This result exhibits convergence only at baryon chemical potentials larger than 2 GeV, which is beyond the density regime reached in physical neutron stars. On the positive side, this implies that one need not worry about the contributions of quark pairing to the perturbative EoS, as they are suppressed by a factor $\Delta^2/\mu_B^2$ with respect to the bulk quantity, where $\Delta\lesssim 100$ MeV is the gap parameter of color superconductivity \cite{Alford:2007xm}. Finally, at small but nozero temperatures, the current state-of-the-art for the unpaired EoS is ${\mathcal O}(g^5)$ \cite{Ipp:2006ij,Kurkela:2016was}.

The range of densities, where the uncertainties of the CET and pQCD EoSs both exceed approx.~20\%, extends from slightly above the saturation density to almost 40$n_s$. In order to describe the EoS inside this no man's land, one clearly needs to relax one of the assumptions made above: either resort to a phenomenological description of the system (see e.g.~\cite{Buballa:2003qv,Kojo:2014rca,Roark:2018uls} and references therein), or deform the theory to allow for an alternative approach via holography \cite{Li:2015uea,Preis:2016fsp,Hoyos:2016zke,Annala:2017tqz,Faedo:2017aoe}. While both of these avenues offer very exciting prospects --- e.g.~through the possibility of discovering properties universal for dense strongly coupled systems \cite{Cherman:2009tw,Hoyos:2016cob,Ecker:2017fyh} --- we shall not dwell on this subject further. Rather, we will now move on to inspect, how the NS matter EoS can be constrained in a model-independent way using only information about the low- and high-density limits discussed above as well as astrophysical observations.

\section{Interpolating from low to high densities}
\label{interpolation}

Keeping in mind the current limitations of ab initio approaches to dense QCD matter discussed above, the most conservative way to estimate the behavior of the NS matter EoS inside the stellar core is to allow the function $\epsilon(p)$ to behave in any physically consistent way between the CET and pQCD regimes. Such an approach has a long history in stellar astrophysics, and in particular extrapolations of low-density nuclear matter EoSs using piecewise polytropes, i.e.~functions exhibiting the generic form $p=\kappa n^\gamma$ in several succeeding subintervals, have been frequently used to extend the applicability of nuclear theory results to higher densities \cite{Hebeler:2013nza}. To consistently quantify one's ignorance about the EoS, it is, however, important to also account for the high-density constraint from pQCD, which was first implemented via a polytropic interpolation in \cite{Kurkela:2014vha}. 

In addition to piecewise polytropic functions, other, more refined ways of carrying out interpolations have been introduced as well, often exhibiting faster convergence towards chosen test functions \cite{Lindblom:2018rfr}. The exact way the interpolation is carried out is, however, in principle unimportant as long as one ensures that the interpolating functions are able to represent any physically meaningful EoS to a sufficient accuracy \cite{Raaijmakers:2018bln}. In \cite{Kurkela:2014vha} and the more recent \cite{Annala:2017llu}, the results of which we will discuss below (see also \cite{Most:2018hfd} for closely related recent work), the interpolation was implemented using 2-, 3- and 4-part polytropes. The parameters of these functions were chosen randomly, ensuring only that the final EoSs meet the following criteria:
\begin{enumerate}
\item Smooth matching of the polytropic EoS to the limiting low- and high-density (CET and pQCD) EoSs.
\item Smooth behavior of the polytropes at each chemical potential where they are matched to each other, with one possible exception corresponding to a first order (deconfinement) phase transition.
\item Subluminality: the speed of sound can never exceed the speed of light.
\end{enumerate}
The most generic set of interpolating 4-tropes satisfying the above constraints can be found from \cite{Annala:2017llu} and is shown in Fig.~1 below. As can be verified from there, the uncertainty in the EoS prior to implementing any astrophysical constraints is sizable, and e.g.~the corresponding radius of a 1.4 solar mass NS varies between 6.6 and 14.7 km.

\section{Astrophysical constraints}
\label{astro}

In order to inspect, how different astrophysical observations constrain the family of NS matter EoSs introduced above, the most straightforward strategy is to analyze the over 200.000 EoSs of \cite{Annala:2017llu} one by one, systematically discarding those that do not meet the desired criteria. The simplest such constraint is clearly related to the existence of the heaviest NSs known to date \cite{Demorest:2010bx,Antoniadis:2013pzd}, which in practice allows us to discard all EoSs whose MR-curves do not reach the two-solar-mass limit.\footnote{Choosing the lower limits of the uncertainty ranges of the two measurements would result in a cut just below $2M_\odot$. As it is, however, highly unlikely that we have observed the most massive NSs in existence, setting the limit at two solar masses should be a safe approximation.} The set of EoSs and corresponding MR-curves that do not fulfill this constraint is denoted by the cyan color in Fig.~1, taken from \cite{Annala:2017llu}. As we readily observe, the EoSs that are ruled out are all very soft at low densities, and would thus predict stars with small radii. Indeed, the two-solar-mass constraint alone allows us to limit the radius of a $1.4M_\odot$ star by 9.9km from below, which further tightens to 10.7km should we only allow for tritropic interpolating functions. Such limits are already very useful, and together with the $2M_\odot$ constraint rule out many model calculations that e.g.~contain hyperons within densities reached in neutron star cores.

\begin{figure}
\begin{center}
\includegraphics[width = 0.48\textwidth]{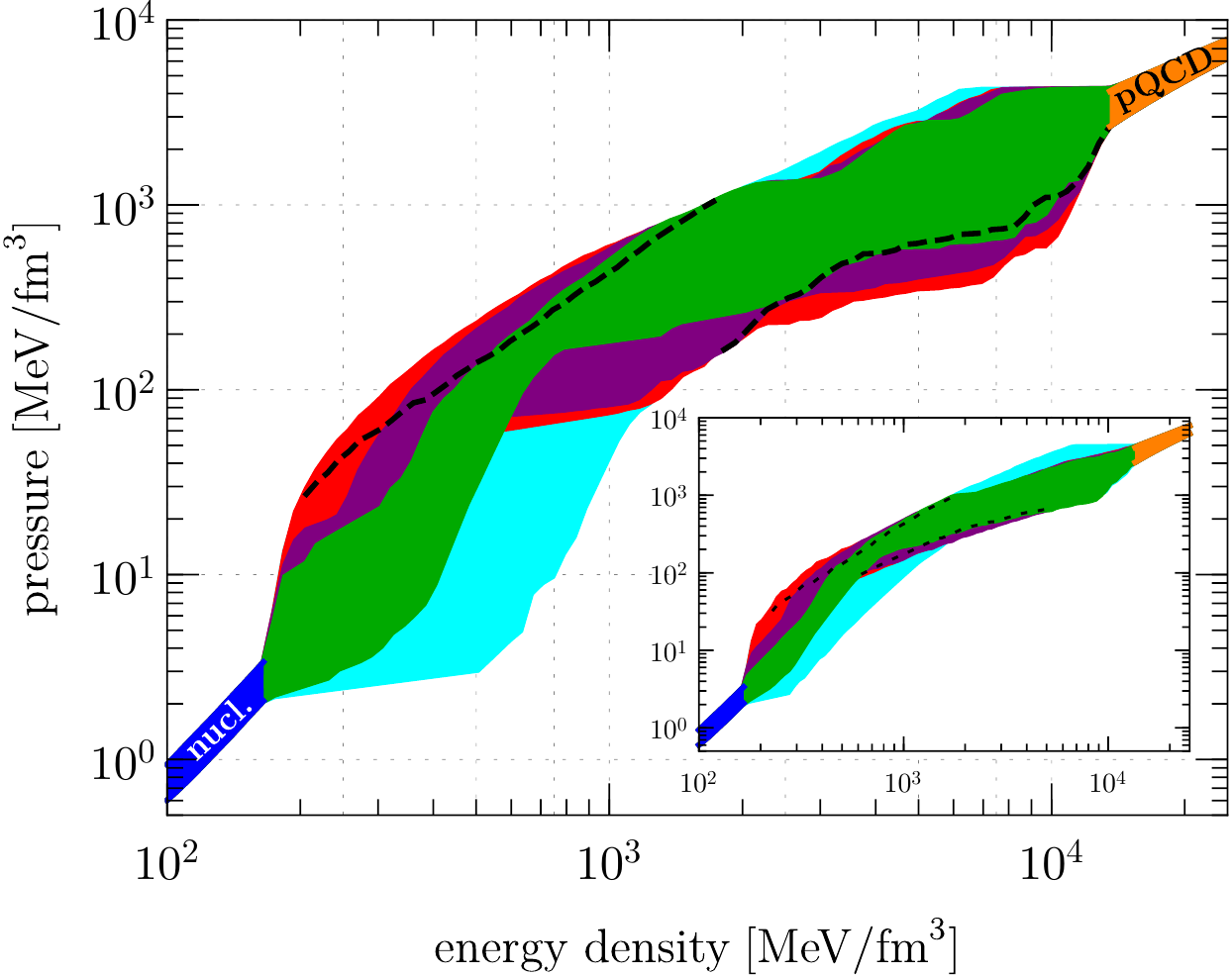} \includegraphics[width=0.48\textwidth]{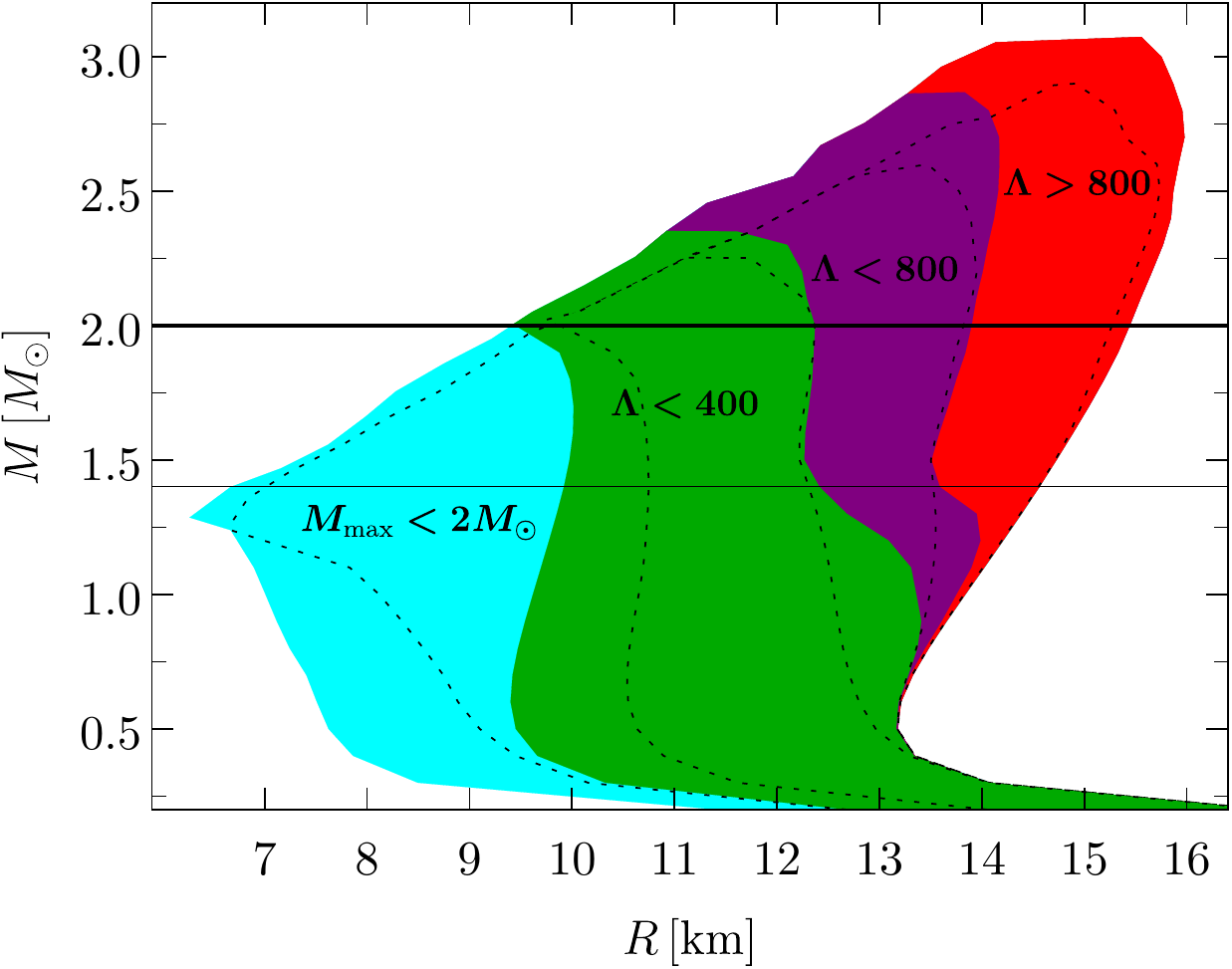}
\caption{
The EoS and MR families obtained in the interpolation study of \cite{Annala:2017llu}. The cyan regions correspond to EoSs that are ruled out by the two-solar-mass constraint, while the red, violet and green areas indicate different tidal deformabilities for $1.4M_\odot$ stars, such that only EoSs producing $\Lambda(1.4M_\odot) < 400$ are shown in green, those with $\Lambda(1.4M_\odot) < 800$ in violet, and all the rest in red. The dashed black lines in the left figure finally show where the lower and upper edges of the EoS band would move if we were to apply the further constraint $M_\text{max} < 2.16M_\odot$ from \cite{Rezzolla:2017aly}, while the inset in the left and the dotted black lines in the right figure correspond to tritropic interpolation.}
\label{MRplot}
\end{center}
\end{figure}

Moving next on to the analysis of GW data, we note that the original Physical Review Letters article of the LIGO and Virgo collaborations \cite{TheLIGOScientific:2017qsa} quoted results for the tidal deformabilities of the two stars in two different ways: firstly as independent measurements, and secondly as an inferred limit for the tidal deformability of a 1.4$M_\odot$ star. Noting that the two comparisons should yield identical results, we concentrate on the latter and categorize our interpolated EoSs based on the $\Lambda$ values they imply for 1.4$M_\odot$ NSs. The original LIGO and Virgo publication \cite{TheLIGOScientific:2017qsa} only placed upper limits for this quantity, suggesting that $\Lambda(1.4M_\odot)<800$ with 90\% credence and $<400$ with 50\% credence. Determining this quantity for each of the EoSs generated in \cite{Annala:2017llu}, we obtain the EoS and MR clouds of Fig.~1, which demonstrates how the tidal deformability constraint efficiently rules out EoSs that are very stiff at low densities, leading to stars with large radii. This is of course understandable considering that the tidal deformability is a measure of a star's susceptibility to tidal forces --- an effect that should correlate with the physical size of the NS in question. Indeed, such a correlation is clearly visible in Fig.~2 of \cite{Annala:2017llu}, where the radii of 1.4$M_\odot$ stars are displayed against the tidal deformabilities of the same configurations for all of the interpolating polytropic EoSs generated in this reference. Finally, the limits one obtains for different quantities characterizing NSs before and after implementing the two-solar-mass and tidal deformability constraints can be found tabulated in Table I of \cite{Annala:2017llu}.

In addition to the indirect information about NS radii that can be obtained from GW measurements, it is of course very interesting to probe this quantity with direct measurements of various kinds. In recent years, promising progress has been achieved in particular in studies of X-ray emission from the NS surface, including most prominently the NICER mission, the results of which are however still to appear \cite{Ozel:2015ykl}. Another interesting line of research concentrates on the analysis of the cooling of thermonuclear X-ray bursts that occur in binary systems, where a NS accretes matter from a companion white dwarf \cite{Nattila:2015jra,Nattila:2017wtj}. A recent result from these studies indicates that the circumferential radius of a NS with a gravitational mass of $1.9\pm 0.3 M_\odot$ is $12.4\pm 0.4$ km at a 68\% credibility level \cite{Nattila:2017wtj}. While this measurement is clearly not at the same level of precision as the ones used in the analyses discussed above, it is tempting to ask, what implications a firm radius measurement of this kind would have, should the mass of the NS in question be accurately known. To this end, we display in Fig.~2 the result of an unpublished study, where it is assumed that on top of the astrophysical constraints discussed above, it has been firmly established that the radius of a $1.9M_\odot$ NS is $12.4\pm0.4$ km. As can be readily witnessed, the effect on the EoS cloud is dramatic in particular at low densities, where a sizable region of soft EoSs now become ruled out. It is straightforward to verify that for such a reduction to take place, it is important that the mass of the NS in question is known accurately, and furthermore very helpful if it is close to the two-solar-mass limit rather than in the $1.4M_\odot$ ballpark. It should additinoally be noted that in this figure, the green region corresponds to $\Lambda(1.4M_\odot)<580$ rather than 400, corresponding to the new 90\% limit of the recent LIGO and Virgo update \cite{Abbott:2018exr}. 

\begin{figure}
\begin{center}
\includegraphics[width=0.48\textwidth]{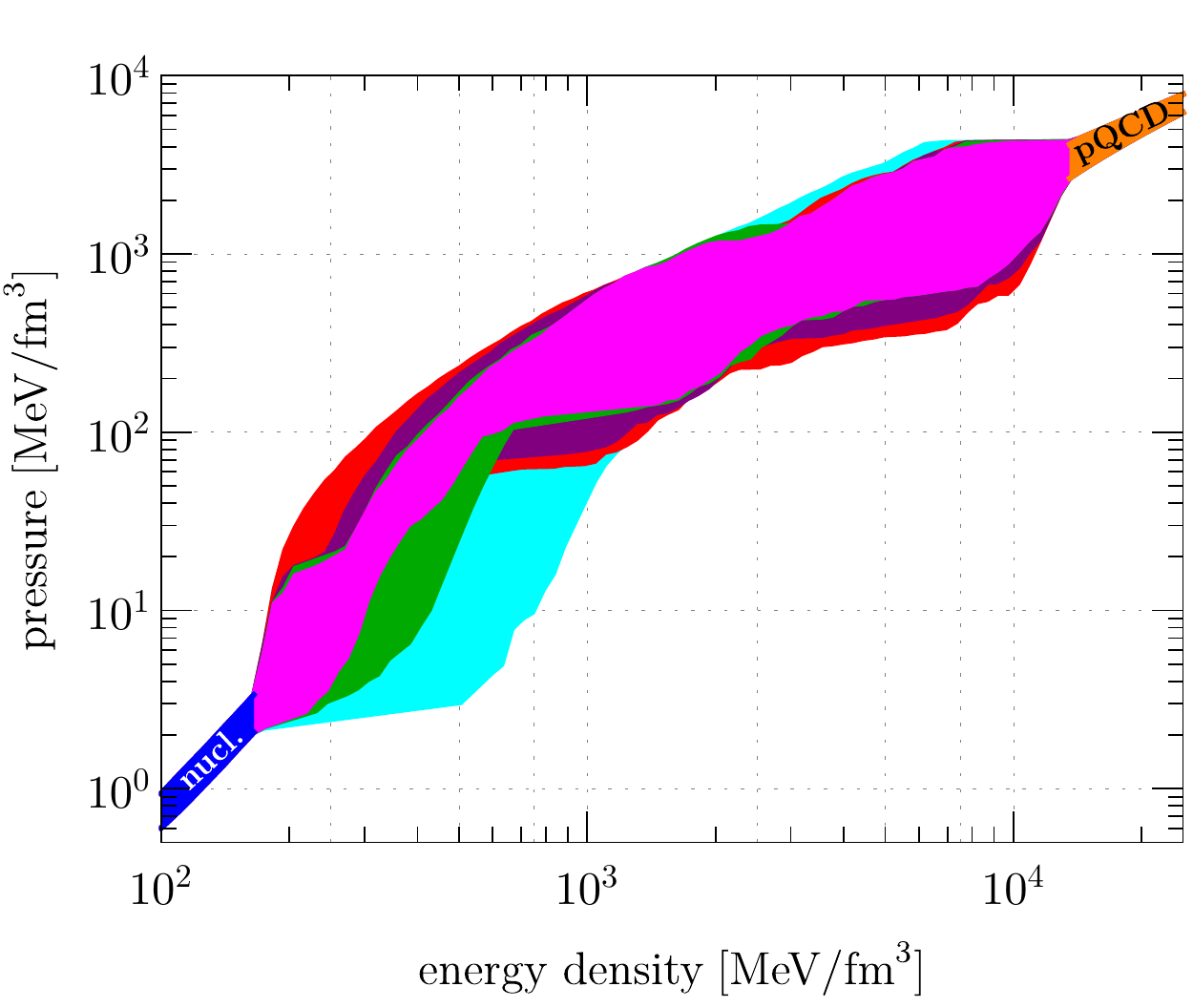}\includegraphics[width=0.48\textwidth]{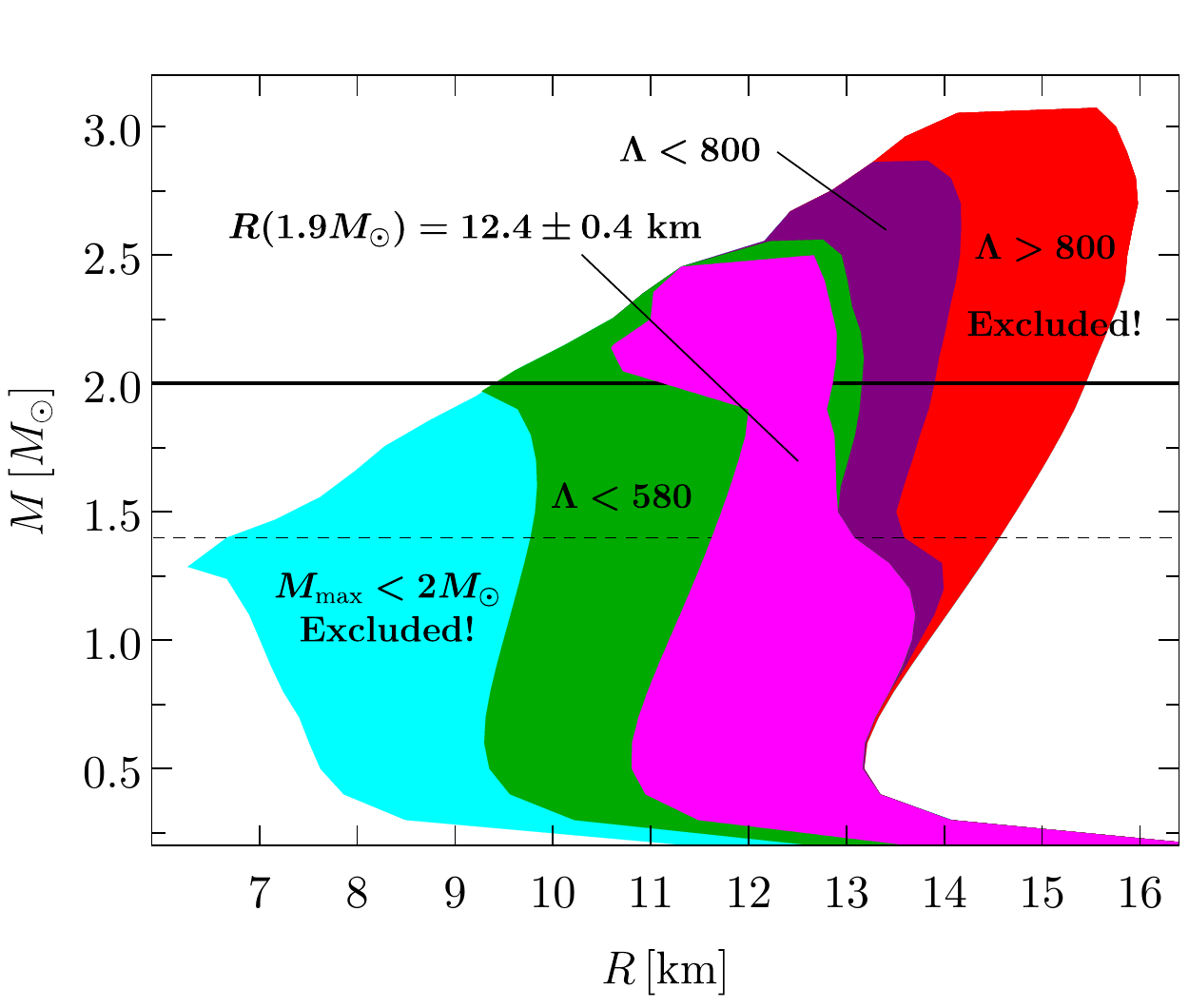}
\caption{EoS and MR families after the newest LIGO and Virgo bound $\Lambda(1.4M_\odot)<580$ (green region) and a hypothetical $R(1.9M_\odot)=12.4\pm0.4$ km radius measurement (pink region) have been taken into account. These figures are based on unpublished work.
}
\label{Lambda_vs_R2}
\end{center}
\end{figure}

\section{Outlook}
\label{outlook}

It is a well-established fact that the inner cores of neutron stars contain QCD matter at densities outside the reach of any terrestrial experiment, making them a potentially very useful laboratory for nuclear physics. It is, however, only in the past decade that measurements of their properties --- in particular the masses and radii of individual stars --- have become precise enough to allow for significant constraints to be set on the most central quantities characterizing NS matter, such as its equation of state. A crucial aid in this process has been the systematic construction of EoS families that fulfill all known consistency criteria, i.e.~possess the correct low- and high-density limits, are thermodynamically consistent and subluminal, and support NSs with properties in agreement with observations (see e.g.~\cite{Hebeler:2013nza,Kurkela:2014vha}). With such a family of EoSs available, it is very straightforward to implement new and tighter observational constraints, including e.g.~the recent LIGO and Virgo limits on the tidal deformabilities of the two stars involved in the event GW170817 \cite{TheLIGOScientific:2017qsa}. It is exactly the result of this kind of an exercise, presented originally in \cite{Annala:2017llu}, that I have discussed in the proceedings contribution at hand.

The first conclusion to be drawn from the results of \cite{Annala:2017llu} is clearly visible in Fig.~1 above: the tidal deformability measurement of LIGO and Virgo acts in practice as a radius constraint, ruling out EoSs that are very stiff at low densities, thereby leading to stars with large radii. This is a valuable result for many reasons, not least because the the GW event GW170817 allowed for a very accurate determination of the so-called chirp mass of the binary system, leading to a radius constraint for a fixed NS mass. This is in contrast with many direct studies of NS radii, where the radius may be measured rather accurately, but the mass of the corresponding star is known with considerably less precision \cite{Nattila:2015jra,Nattila:2017wtj}.

Considering that we have only witnessed one NS merger so far --- and that considerable progress can be expected from individual mass and radius measurements as well --- the prospects in the quest to determine the NS matter EoS look very bright. With more observational data, it can be expected that calculations of the type presented above will become sensitive to very nontrivial details of the underlying microphysics, including traces of a possible first order phase transition in the EoS. This is, however, only one part in the challenge to understand the nature of the dense QCD matter inside NS cores: to arrive at the right physics conclusions concerning e.g.~the phases of QCD matter realized in these systems, one must simultaneouly vigorously develop the ways, in which dense nuclear and quark matter are currently described. This is currently one of the most pressing challenges of modern QCD theory, and likely continues to be that for the foreseeable future.

\section*{Acknowledgments}
The work of AV is supported by the Academy of Finland, grant no.~1303622, as well as by the European Research Council, grant no.~725369.








\end{document}